\documentclass[11pt,twoside]{article}

\usepackage{asp2004}
\usepackage{epsf}
\usepackage{psfig}
\usepackage{lscape}
\usepackage{graphicx}

\begin{document}
\title{Small Magellanic Cloud Be stars: colour-magnitude relations and mass-loss}    
\author{W.J. de Wit}
\affil{Laboratoire d'Astrophysique de Grenoble, Universit\'{e} Joseph Fourier, BP 53, 38041 Grenoble Cedex 9, France (dewit@obs.ujf-grenoble.fr) }
\author{H.J.G.L.M. Lamers}
\affil{Astronomical Institute and SRON Laboratory for Space Research,
Utrecht University, Princetonplein 5, 3584CC Utrecht, The Netherlands}
\author{ J.B. Marquette, J.P. Beaulieu}
\affil{IAP, 98$^{\rm bis}$ Boulevard Arago, 75014 Paris, France}

\begin{abstract} 
  We present an analysis of optical lightcurves of Small Magellanic Cloud (SMC) Be-type
  stars. Observations show that (1) optical excess flux is correlated with near-IR excess
  flux indicating a similar mechanism and (2) the lightcurves can trace out
  ``loops'' in a colour-magnitude diagram. A simple model for the time dependence
  of bound-free and free-free (bf-ff) emission produced by an outflowing circumstellar
  disk gives reasonable fits to the observations.
\end{abstract}
\noindent
Mennickent et al. (2002) present and describe optical OGLE lightcurves for a
large number of blue stars in the SMC. A subset of these stars is found to be
irregular variable on timescales of weeks to months and amplitudes up to $1^{\rm
m}$. An example is given in the left panel of Fig.\,1. A cross-correlation with
2MASS shows that there exists a correlation between same epoch optical excess
and near-IR excess (Fig.\,1, right panel). The optical excess at 2MASS epoch is with respect
to the brightness minimum, which should then correspond to the stellar
photosphere. To assure this, we plot only those stars that have actual B-type
optical colours at brightness minimum.

Some 40\% of our sample stars that have an amplitude of variation $>0.2^{m}$
(101 objects) trace out a ``loop'' in a colour-magnitude diagram (CMD). A good
example is given in Fig.\,2, other examples can be found in de Wit et
al. (2006)\nocite{}. It is noteworthy that for 90\% the loops are traced out in a
clock-wise sense. Only a small minority makes the loop in an anti clock-wise
sense.
\begin{figure*}[!ht]
  \begin{center}
    \includegraphics[width=5cm,height=6cm,angle=90]{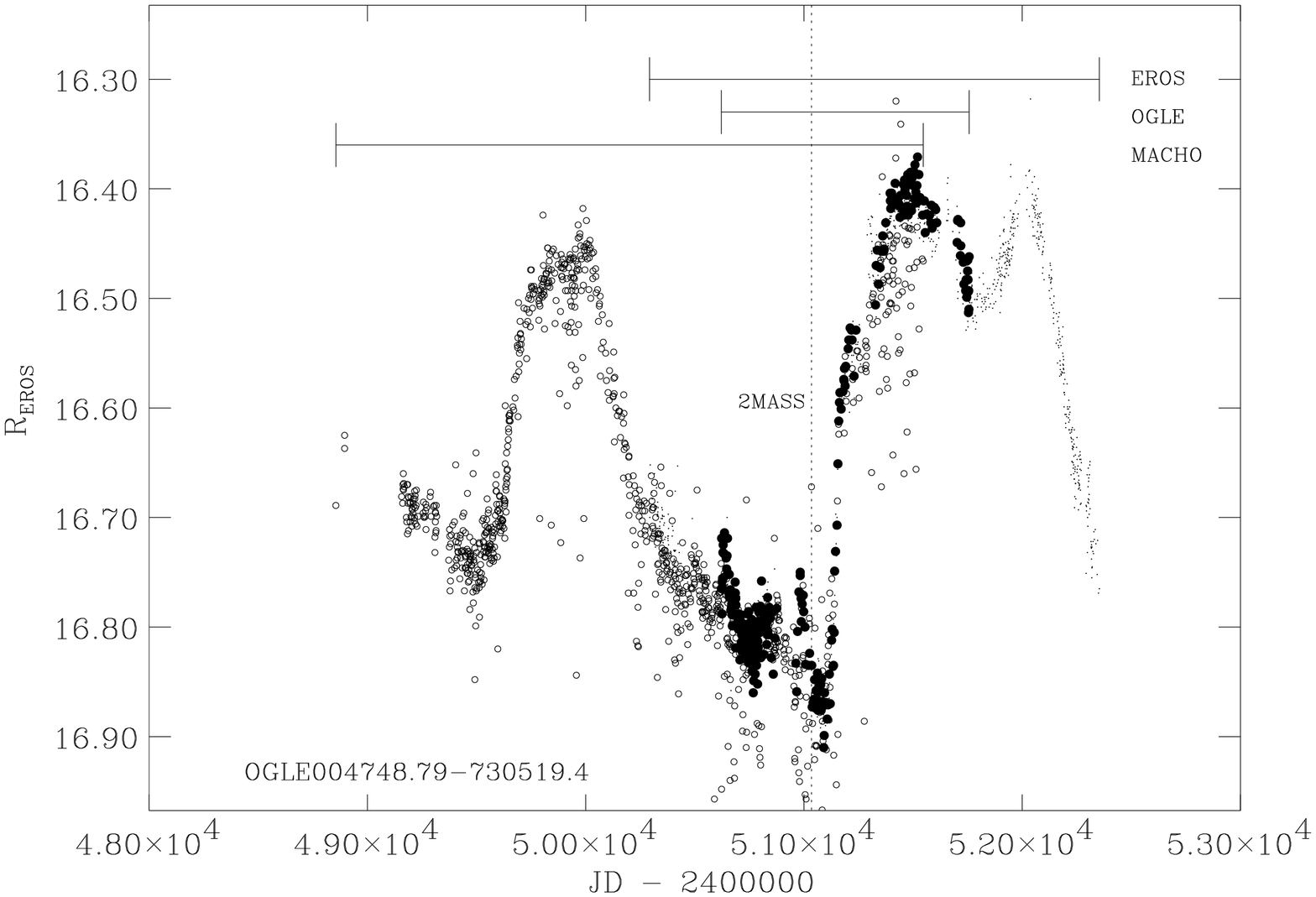}
    \includegraphics[width=5cm,height=6cm,angle=90]{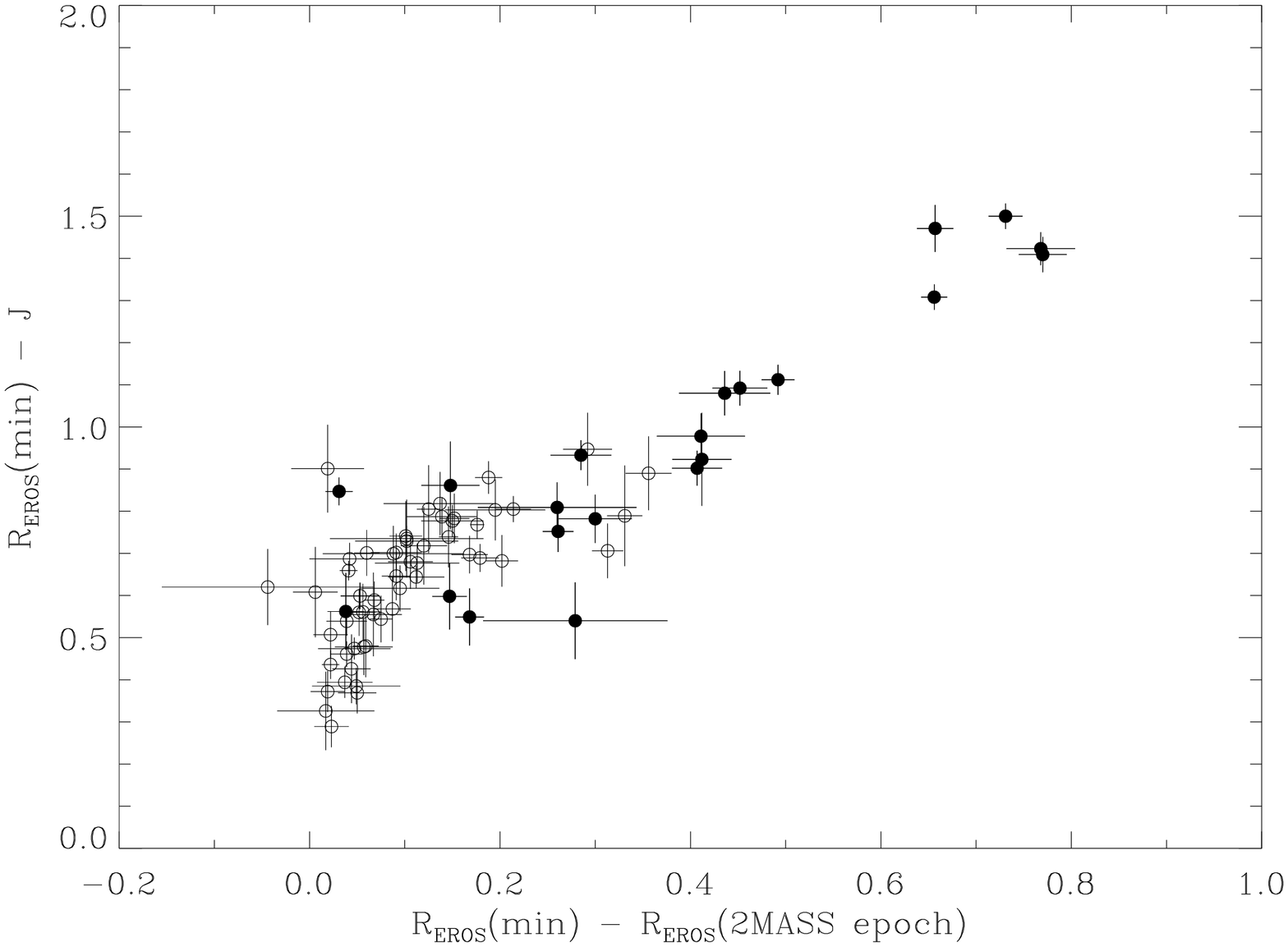}
  \end{center}
  \caption{
    {\itshape Left:\/} Optical variability of a blue SMC star. Measurements are from
    the EROS, OGLE, and MACHO microlensing experiments.
    {\itshape Right:\/} Correlation between near-IR flux excess (from 2MASS) and
    optical flux excess (from EROS light curve) at the same epoch. 
    \label{fig:1}}
\end{figure*}
\begin{figure*}[!ht]
  \begin{center}
    \includegraphics[width=5cm,height=12.5cm,angle=90]{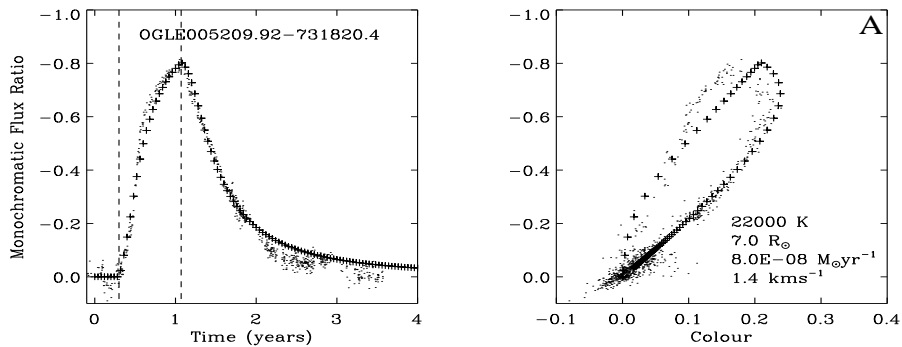}
  \end{center}
  \caption{Observed (small dots) and modeled (crosses) variability. Model
    parameters are indicated, these values are constants with radius.
    {\itshape Left:\/}  Light curve, the Be star loses mass during the period
    indicated by the vertical dashed lines.
    {\itshape Right:\/} Corresponding CMD, notice the ``loop'', in this case clockwise.
    \label{fig:2}}
\end{figure*}

We develop a simple model to describe the observed variability. The model
consists of a mass losing star and an outflowing, geometrically thin CS disk
producing bf-ff emission. The disk is described by powerlaws in temperature,
opening angle and outflow velocity. For an exact model description, see de Wit
et al. (2006). Using this model, one can interpret the observed CMD behaviour in
the following way. Once the B star starts losing mass a CS disk is built up. At
small radii the disk is optically thick, and is the dominant source of the total
flux (and colour) excesses. Once the mass-loss halts, the optically thick part
of the disk is removed (this depends on wavelength) and the predominantly
optically thick disk makes the transition to an optically thin outflowing
ring. Observationally, this results in a clock-wise traversed loop in a
CMD. Counter clock-wise loops are produced when matter flows in, rather than
out. As mentioned above, inflow is observed in only a small minority of cases.


\acknowledgements 
I wish to express my deepest gratitudes to Henny for nearly a decade of priceless
support. Happy 65th birthday!


\end{document}